\documentclass[12pt,onecolumn,prc,showpacs,preprintnumber,amsmath,
floatfix,amssymb]{revtex4}
\usepackage{epsfig}
\usepackage{graphicx}% Include figure files
\usepackage{dcolumn}% Align table columns on decimal point
\usepackage{bm}% bold math
\def\r{\mbox{{\bf  r}}}
\def\p{\mbox{\boldmath $p$}}
\def\q{\mbox{\boldmath $q$}}
\def\k{\mbox{\boldmath $k$}}
\def\t{\mbox{\boldmath $t$}}
\allowdisplaybreaks[4]

\begin{document}
%\preprint{  }
\title{Analysis of flux-integrated cross sections for quasi-elastic neutrino 
charged-current scattering off $^{12}$C at MiniBooNE energies.}
\author{A.~V.~Butkevich}
\affiliation{ Institute for Nuclear Research,
Russian Academy of Sciences,
60th October Anniversary Prosp. 7A,
Moscow 117312, Russia}
\date{\today}% It is always \today, today,
\begin{abstract}

Flux-averaged and flux-integrated cross sections for quasi-elastic neutrino 
charged-current scattering on nucleus are analyzed. It is shown that the 
flux-integrated differential cross sections are nuclear model-independent. We 
calculate these cross sections using the relativistic distorted-wave impulse 
approximation and relativistic Fermi gas model with the Booster Neutrino 
Beamline flux and compare results with the recent MiniBooNE experiment data. 
Within these models an axial mass $M_A$ is extracted from a fit of the 
measured $d\sigma/dQ^2$ cross section. The extracted value of $M_A$ is 
consistent with the MiniBooNE result. The measured and calculated double 
differential cross sections $d\sigma/dTd\cos\theta$ generally agree within the 
error of the experiment. But the Fermi gas model predictions are completely 
off of the data in the region of low muon energies and scattering angles.  
\end{abstract}

\maketitle

\section{Introduction}

The current~\cite{MiniB1, MINOS1, OPERA, T2K} and planed~\cite{NOvA} set of 
accelerator-based neutrino experiments use extremely intense neutrino 
beamlines for precise measurements of the observed neutrino mass splitting and 
mixing angles and detailed experimental study of the neutrino mixing matrix. 
The data of these experiments will greatly extend the statistics due to
extremely intense neutrino beamline. 

In this situation, the statistical uncertainties should be negligible as 
compared to systematic errors. An important source of systematic uncertainties 
is related to the neutrino-nucleus $(\nu A)$ cross sections. The neutrino 
beams of high intensity cover the few-GeV energy range. In this energy regime,
 the dominant contribution to $\nu A$ cross section comes from charged-current 
(CC) quasielastic (QE) scattering and resonance production processes. 
In the long-base line neutrino oscillation experiments near and far detectors
 are used to normalize the neutrino flux at production and to search for the 
neutrino oscillation effects. While many unknown quantities are eliminated 
in these experiments by considering ratios of far to near events, the 
cancellation is not complete due to differences in neutrino flux and 
backgrounds in the near and far detectors. Thus, in order to permit precision 
oscillation measurements it is important to have an accurate characterization 
of the CCQE differential cross sections over wide span neutrino energies.

The current data on CCQE scattering come from a variety of experiment 
operating at differing energies and with different nuclei. The existing data on
 (anti)neutrino CCQE scattering come mostly from bubble chamber experiments, 
which suffer from small statistic. In general, the experimental execution and 
data interpretation are non-trivial for several reason. Neutrino beams 
typically span a wide energy range. The neutrino flux itself is often poorly 
known and a background from resonance processes are frequently significant and 
it is difficult for separating from the CCQE signal. Therefore the total QE 
cross sections measured in different experiments with accuracy of 20-40\% and 
even within such large uncertainties some results contradict each other. The 
difference between the total quasielastic cross sections, calculated within the 
framework of various models~\cite{Moniz, Kim1, Nieves, Meucci1, Maieron, 
Martinez, Kim2, Athar, Leither, Martini, BAV1, BAV2, Boyd} is lower than the 
spread in data. 

More information about neutrino-nuclear CCQE interaction can be obtained from 
the analysis of the charged-current QE event 
distributions and $d\sigma/d Q^2$ differential cross sections as functions of 
$Q^2$ (squared four-momentum transfer)~\cite{Kuzmin}. The shape of these 
distributions is sensitive to the $Q^2$ dependence of two vector, 
$F_{1,2}(Q^{2})$, one axial-vector $F_A(Q^2)$ form factors and nuclear effects. 
The vector form factors are well-known from electron scattering. For the 
axial-vector form factors the dipole parametrization with one free parameter 
$M_A$ (axial mass) is mainly used. This parameter controls the $Q^2$ 
dependence of $F_A(Q^2)$, and ultimately, the normalization of the predicted 
cross sections. The dipole parametrization has no strict theoretical basis and 
the choice of this parametrization is made by the analogy with 
electroproduction. To describe the nuclear effects, neutrino CCQE models 
typically employ a relativistic Fermi gas model (RFGM)~\cite{Moniz} in which 
the nucleons with a flat nucleon momentum distribution up to the same Fermi 
momentum $p_F$ and nuclear binding energy $\epsilon_b$. 
The experimental values of $M_A$ extract from the (anti)neutrino CCQE 
scattering data, i.e. from the analysis of the shape of the 
$Q^2$- distributions and from the direct measurements of the total cross 
sections. They show very wide spread from roughly 0.7 to 1.2 GeV and the 
resulting world-average $M_A=1.03\pm0.02$ GeV~\cite{Bernard}. 

Several experiments have recently reported new results on CCQE scattering from 
high-statistics data samples with intense, well-understood neutrino beams. 
The NOMAD experiment~\cite{NOMAD} observe an $M_A$ value and cross section 
(from data taken on carbon) consistent with prior world-average. However, data 
of~\cite{MiniB1, K2K1, K2K2, SciB} and~\cite{MINOS2} (preliminary result), 
collected on carbon, oxygen, and iron targets, have indicated a somewhat 
larger value for $M_A$ (by $\approx 10-30\%$). In these experiments the shape 
of the $Q^2$-distribution was analyzed.

This data show a disagreement with the RFGM predictions. The data samples 
exhibit deficit in the region of low $Q^2\leq 0.2$ (GeV/c)$^2$ (so-called low-
$Q^2$ problem). As it is known the comparison with the low-energy QE 
electron-nucleus scattering data, the RFGM description of this region is 
not accurate enough~\cite{BAV3}. In the region of high-$Q^2$ the data excess 
is observed, and value of $M_A$, obtained from a fit to the measured data,
 is higher than the results of the previous experiments. The collection of 
existing results remains puzzling. The next experiments MINERvA~\cite{MINERvA} 
and MicroBooNE~\cite{MicroB} as well as T2K~\cite{T2K} and 
NOvA~\cite{NOvA} near detectors will be able to make more precise 
measurements of the CCQE cross sections in a wide range of energies 
and for various nuclear targets.  

The uncertainties in the theoretical description 
of the quasielastic neutrino-nucleus scattering could be considerably reduced 
if new model-independent absolute differential cross section could be provided.
 The first measurement of the flux-integrated double-differential cross 
section (in muon energy and angle) for CCQE scattering on carbon has been 
produced in MiniBooNE experiment~\cite{MiniB2}. This cross section contains 
the most complete and model-independent information that is available from 
MiniBooNE for the CCQE process. 

The aim of this work is to test the RFGM and relativistic distorted-wave 
impulse approximation (RDWIA) predictions against the MiniBooNE data
~\cite{MiniB2}. In the framework of these approaches we extract the values of 
axial mass from the measured flux-integrated $d\sigma/d Q^2$ cross section. 
Then, we calculate with extracted values of $M_A$ the flux-integrated 
differential and flux-unfolded total cross sections and compare the results 
with data.

The outline of this paper is the following. In Sec. II we present briefly the
RDWIA model and discuss the flux-averaged and flux-integrated differential 
cross sections. The results are presented in Sec. III. Our conclusions are 
summarized in Sec. IV. 

\section{Model, flux-averaged and flux-integrated differential cross 
sections}

We consider neutrino charged-current QE exclusive
\begin{equation}\label{qe:excl}
\nu(k_i) + A(p_A)  \rightarrow \mu(k_f) + N(p_x) + B(p_B),      %1
\end{equation}
and inclusive
\begin{equation}\label{qe:incl}
\nu(k_i) + A(p_A)  \rightarrow \mu(k_f) + X                      %2
\end{equation}
scattering off nuclei in the one-W-boson exchange approximation. Here 
$k_i=(\varepsilon_i,\k_i)$ 
and $k_f=(\varepsilon_f,\k_f)$ are the initial and final lepton 
momenta, $p_A=(\varepsilon_A,\p_A)$, and $p_B=(\varepsilon_B,\p_B)$ are 
the initial and final target momenta, $p_x=(\varepsilon_x,\p_x)$ is the 
ejectile nucleon momentum, $q=(\omega,\q)$ is the momentum transfer carried by 
the virtual W-boson, and $Q^2=-q^2=\q^2-\omega^2$ is the W-boson 
virtually. 

\subsection{Model}

The formalism of charged-current QE exclusive and inclusive reactions is 
described in~\cite{BAV1}. All the nuclear structure information and final 
state interaction effects (FSI) are contained in the weak CC nuclear tensors 
$W_{\mu \nu}$, which are given by bilinear product of the transition matrix 
elements of the nuclear CC operator $J_{\mu}$ between the initial nucleus 
state $|A\rangle$ and the final state $|B_f\rangle$ as 
\begin{eqnarray}
\label{Eq3}
W_{\mu \nu } &=& \sum_f \langle B_f,p_x\vert                           %8
J_{\mu}\vert A\rangle \langle A\vert
J^{\dagger}_{\nu}\vert B_f,p_x\rangle,              %%%  \nn
\label{W}
\end{eqnarray}
where the sum is taken over undetected states.

We describe CCQE neutrino-nuclear scattering in the impulse approximation (IA),
 assuming that the incoming neutrino interacts with only one nucleon, which is 
subsequently emitted, while the remaining (A-1) nucleons in the target are 
spectators. The nuclear current is written as the sum of single-nucleon 
currents. Then, the nuclear matrix element in Eq.(\ref{Eq3}) takes the form
\begin{eqnarray}\label{Eq4}
\langle p,B\vert J^{\mu}\vert A\rangle &=& \int d^3r~ \exp(i\t\cdot\r)
\overline{\Psi}^{(-)}(\p,\r)
\Gamma^{\mu}\Phi(\r),                                                     %12
\end{eqnarray}
where $\Gamma^{\mu}$ is the vertex function, $\t=\varepsilon_B\q/W$ is the
recoil-corrected momentum transfer, $W=\sqrt{(m_A+\omega)^2-\q^2}$ is the
invariant mass, $\Phi$ and $\Psi^{(-)}$ are relativistic bound-state and
outgoing wave functions.

The single-nucleon charged current has $V{-}A$ structure $J^{\mu} = 
J^{\mu}_V + J^{\mu}_A$. For a free-nucleon vertex function 
$\Gamma^{\mu} = \Gamma^{\mu}_V + \Gamma^{\mu}_A$ we use the CC2 vector 
current vertex function $\Gamma^{\mu}_V$. The weak vector form factors are 
related to the corresponding electromagnetic ones for protons and neutrons by 
the hypothesis of the conserved vector current. We use the approximation 
of Ref.~\cite{MMD} on the nucleon form factors.
Because the bound nucleons are off shell we employ the de Forest 
prescription~\cite{deFor} and Coulomb gauge for off-shell vector current 
vertex $\Gamma^{\mu}_V$. The vector-axial and pseudoscalar form factors are 
parametrized as a dipole with the axial-vector mass, which controls the $Q^2$ 
dependence of $F_A$, and ultimately, the normalization of the predicted cross 
section. 

According to the JLab data~\cite{Dutta, Kelly1} the occupancy of the 
independent particle shell-model (IPSM) orbitals of ${}^{12}$C equals on 
average 89\%. In this work we assume that the missing strength (11\%) can be 
attributed to the short-range nucleon-nucleon ($NN$) correlations in the 
ground state, leading to the appearance of the high-momentum (HM) and 
high-energy component in the nucleon distribution in the target. To estimate 
this effect in the inclusive cross sections we consider a phenomenological 
model which incorporates both the single-particle nature of the nucleon 
spectrum at low energy (IPSM orbitals) and the high-energy and high-momentum 
components due to $NN$ correlations.

In the independent particle shall model the relativistic wave functions of the 
bound nucleon states $\Phi$ are obtained as the self-consistent (Hartree--
Bogoliubov) solutions of a Dirac equation, derived, within a relativistic mean 
field approach, from Lagrangian containing $\sigma, \omega$, and $\rho$ mesons 
(the $\sigma-\omega$ model)\cite{Serot}. 
We use the nucleon bound-state functions calculated for carbon by the TIMORA 
code~\cite{Horow2} with the normalization factors $S(\alpha)$ relative to the 
full occupancy of the IPSM orbitals of ${}^{12}$C: $S(1p_{3/2})$=84\%, 
$S(1s_{1/2})$=100\%, and an average factor of about 89\%. These estimations of
the depletion of hole states follow from the RDWIA analysis of 
${}^{12}$C$({e},e^{\prime}{p})$ for $Q^2 < 2$ (GeV/c)$^2$~\cite{Kelly1}
and are consistent with a direct measurement of the spectral function using 
${}^{12}$C$({e},e^{\prime}{p})$ in parallel kinematics~\cite{Rohe}, which 
observed approximately 0.6 protons in a region attributable to a 
single-nucleon knockout fromcorrelated cluster. 
%%%

For the outgoing nucleon the simplest choice is to use plane-wave function 
$\Psi$, i.e., no interactions is considered between the ejected nucleon $N$ 
and the residual nucleus B (PWIA - plane-wave impulse approximation). For a 
more realistic description, FSI effects should be taken into account. In the 
RDWIA the distorted-wave function $\Psi$ are evaluated as solution of a Dirac 
equation containing a phenomenological relativistic optical potential. The 
channel coupling in the FSI~\cite{Kelly2} of the $N+B$ system is taken into 
account. The relativistic optical potential consists of a real part which 
describes the rescattering of the ejected nucleon and of an imaginary part that 
accounts for absorption of it into unobserved channels.

Using the direct Pauli reduction method the system of two coupled first-order 
radial Dirac equations can be reduced to a single second-order 
Schr\"odinger-like equation for the upper component of Dirac wave function 
$\Psi$. We use the LEA program~\cite{Kelly3} for the numerical calculation of 
thedistorted wave functions with the EDAD1 parametrization~\cite{Cooper} of the 
relativistic optical potential for carbon. This code, initially designed for 
computing exclusive proton-nucleus and electron-nucleus scattering, was 
successfully tested against A$(e,e'p)$ data~\cite{Fissum, Dutta} and we adopted 
this program for neutrino reactions.  

A complex optical potential with a nonzero imaginary part generally produces 
an absorption of the flux. For the exclusive A$(l,l'N)$ channel this reflects 
the coupling between different open reaction channels. However, for the 
inclusive reaction, the total flux must conserve. In Ref.
\cite{Meucci1, Meucci3} it was shown that the inclusive CCQE neutrino cross 
section of the exclusive channel A$(l,l'N)$, calculated with only the real 
part of the optical potential is almost identical to those of the Green's 
function approach~\cite{Meucci1, Meucci2} in which the FSI effects on inclusive 
reaction A$(l,l'X)$ is treated by means of a complex potential and the total 
flux is conserved. We calculate the inclusive and total cross sections with 
the EDAD1 relativistic optical potential in which only the real part is 
included.

The inclusive cross sections with the FSI effects in the presence of the 
short-range $NN$ correlations were calculated using the method proposed in
 Ref.~\cite{BAV1}. In this approach the contribution of the $NN$ correlated 
pairs is evaluated in the PWIA model. We use the general expression for the 
high-momentum and high-energy part of the spectral function from Ref.
~\cite{Kulagin} with the parametrization for the nucleon high-momentum 
distribution from Ref.~\cite{Atti}, which was renormalized to value of 11\%.
The FSI effects for the high-momentum 
component is estimated by scaling the PWIA cross section with $\Lambda
(\varepsilon_f\Omega_f)$ function determined in Ref.~\cite{BAV1}.

\subsection{Flux-averaged and flux-integrated differential cross sections}

In neutrino experiments the differential cross sections of CCQE 
neutrino-nucleus scattering are measured within rather wide ranges of the 
(anti)neutrino energy spectrum. Therefore flux-averaged and flux-integrated 
differential cross sections can be extracted.

\begin{figure*}
  \begin{center}
    \includegraphics[height=16cm,width=16cm]{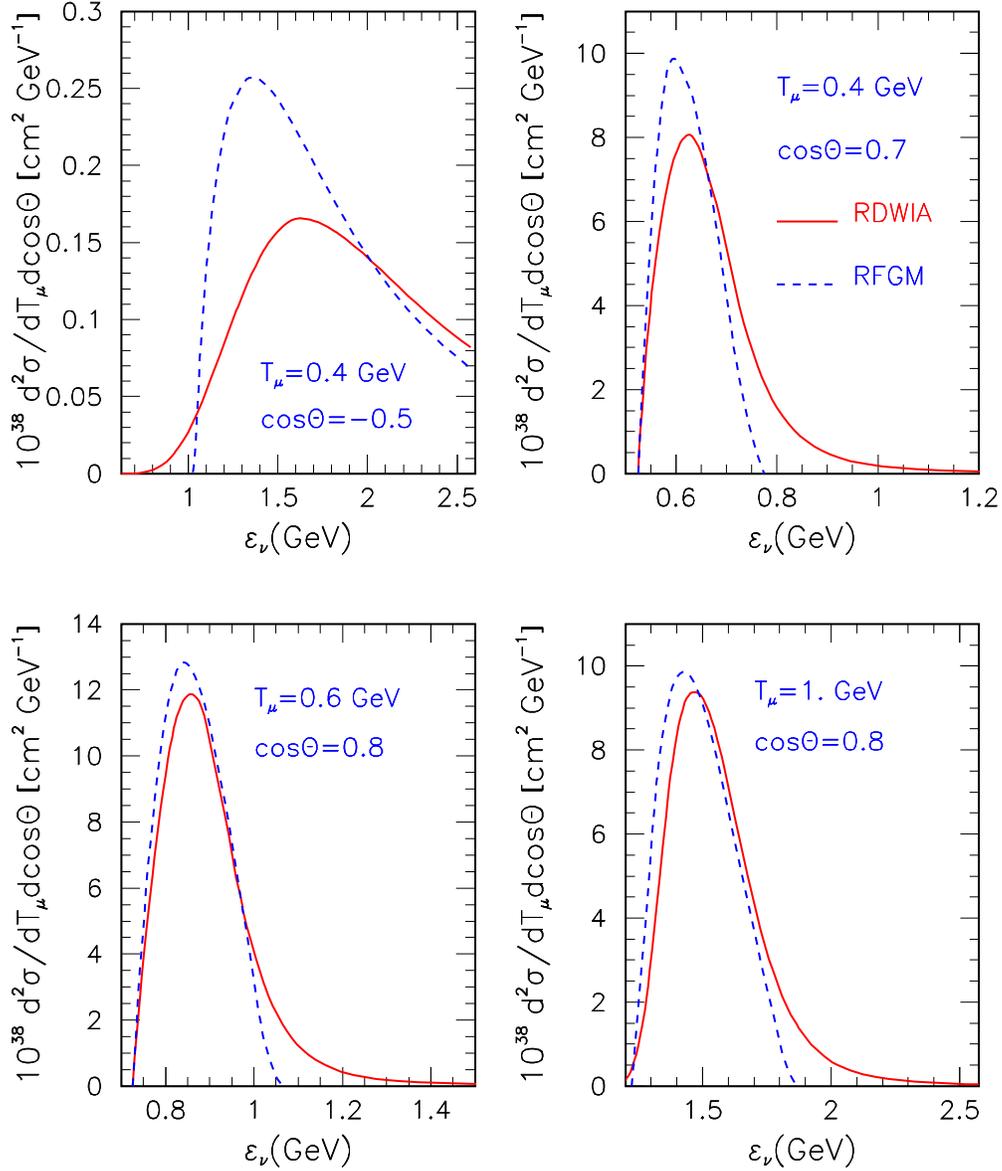}
  \end{center}
  \caption{\label{Fig1}(Color online) Double differential cross sections vs the 
neutrino energy calculated in the RDWIA (solid line) and RFGM (dashed line) 
approaches for the four values of $(T,\cos\theta)$:(0.4 GeV, -0.5), 
(0.4 GeV, 0.7), (0.6 GeV, 0.8), and (1 GeV, 0.8).} 
\end{figure*}

Because the $\nu_{\mu}$- mode of beams incorporates $\nu_{\mu}$ and 
$\bar{\nu}_{\mu}$ spectra the flux-averaged double differential cross section 
$\sigma/dT d\cos\theta$ in muon kinetic energy $T$ and muon scattering angle 
$\theta$ is the sum of neutrino and antineutrino cross sections
\begin{eqnarray}
\label{Eq5}
\left\langle \frac{d^2\sigma}{dTd\cos\theta}\right\rangle &=& 
\left\langle \frac{d^2\sigma^{\nu}}{dTd\cos\theta}\right\rangle +
\left\langle \frac{d^2\sigma^{\bar{\nu}}}{dTd\cos\theta}\right\rangle,
\end{eqnarray}
where
%%%%%%%%%%%%
\begin{eqnarray}
\label{Eq6}
\left\langle \frac{d^2\sigma^{\nu,~\bar{\nu}}}{dTd\cos\theta}(T,\cos\theta)
\right\rangle &=& 
\int_{\varepsilon_1}^{\varepsilon_2}W_{\nu,~\bar{\nu}}(T,\cos\theta,\varepsilon_i)
\frac{d^2\sigma^{\nu,~\bar{\nu}}}{dTd\cos\theta}(T,\cos\theta,\varepsilon_i) 
d\varepsilon_i,
\end{eqnarray}
%%%%%%%%%
and $W_{\nu,~\bar{\nu}}$ are weight functions. The normalization of these 
functions is  given by
%%%%%%%%%%%%
\begin{eqnarray}
\label{Eq7}
\int_{\varepsilon_1}^{\varepsilon_2}[W_{\nu}(T,\cos\theta,\varepsilon_i) + 
W_{\bar{\nu}}(T,\cos\theta,\varepsilon_i)]d\varepsilon_i &=& 1.
\end{eqnarray}
%%%%%%%%%
The weight functions are defined as follows
%%%%%%%%%%%%
\begin{eqnarray}
\label{Eq8}
W_{\nu,~\bar{\nu}}(T,\cos\theta,\varepsilon_i) &=& I_{\nu~\bar{\nu}}(\varepsilon_i)/
\Phi(T,\cos\theta),
\end{eqnarray}
%%%%%%%%%
where $I_{\nu,~\bar{\nu}}$ is the neutrino (antineutrino) spectrum in $\nu$-mode 
of the flux and
%%%%%%%%%%%%
\begin{eqnarray}
\label{Eq9}
\Phi(T,\cos\theta) &=& \int_{\varepsilon_1}^{\varepsilon_2}[I_{\nu}(\varepsilon_i)+
I_{\bar{\nu}}(\varepsilon)]d\varepsilon_i
\end{eqnarray}
%%%%%%%%%
is the neutrino and antineutrino flux which give the contribution to the 
measured double differential cross section at the fixed values of 
$(T,\cos\theta)$. This flux depends on $(T,\cos\theta)$ due to the limits 
of integration in Eqs.~\eqref{Eq6},~\eqref{Eq7} and~\eqref{Eq9} which are 
functions of $(T,\cos\theta)$, i.e. 
$\varepsilon_i=\varepsilon_{\min}(T,\cos\theta)$ and 
$\varepsilon_2=\varepsilon_{\max}(T,\cos\theta)$. In Fig.~\ref{Fig1} the double 
differential cross sections, calculates within the RDWIA and RFGM (with the 
Fermi momentum $p_F=221$ MeV/c and a binding energy $\epsilon_b=25$ MeV for 
carbon) are shown as functions of neutrino energy. Apparently the ranges 
$[\varepsilon_{max}(T,\cos\theta) -\varepsilon_{min}(T,\cos\theta)]$ where 
$d\sigma^2/dTd\cos\theta$ is not equals to zero are different in the RDWIA and 
RFGM. 
Therefore the value of $\Phi(T,\cos\theta)$ is model-dependent and ultimately 
the weight functions and the cross section 
$\langle d^2\sigma/dT d\cos\theta\rangle$ depend on nuclear models too. Note 
that the flux $\Phi(T,\cos\theta)$ should be used to extract the measured 
flux-averaged cross section in the {\it i,j}-bins of $(T,\cos\theta)$ 
variables (for example see Eq.(3) in Ref.~\cite{MiniB2}).

Similarly, the flux-averaged $d\sigma/dQ^2$ cross section can be written as 
sum
\begin{eqnarray}
\label{Eq10}
\left\langle \frac{d\sigma}{dQ^2}\right\rangle &=& 
\left\langle \frac{d\sigma^{\nu}}{dQ^2}\right\rangle +
\left\langle \frac{d\sigma^{\bar{\nu}}}{dQ^2}\right\rangle,
\end{eqnarray}
where
%%%%%%%%%%%%
\begin{eqnarray}
\label{Eq11}
\left\langle \frac{d\sigma^{\nu,~\bar{\nu}}}{dQ^2}(Q^2,T_{th})\right\rangle &=& 
\int_{\varepsilon_1}^{\varepsilon_2}W_{\nu,~\bar{\nu}}(Q^2,\varepsilon_i)
\frac{d\sigma^{\nu,~\bar{\nu}}}{dQ^2}(Q^2,T_{th},\varepsilon_i) 
d\varepsilon_i,
\end{eqnarray}
%%%%%%%%%
and $T_{th}$ is the muon threshold energy after all cuts for CCQE events 
selection.
% FIGURE 1
%%%%%%%%%%%%%%%%%%%%%%%%%%%%%%%%%%%%
The weight functions in Eq.~\eqref{Eq11} are defined as follows
%%%%%%%%%%%%
\begin{eqnarray}
\label{Eq12}
W_{\nu,~\bar{\nu}}(Q^2,\varepsilon_i) &=& I_{\nu,~\bar{\nu}}(\varepsilon_i)/
\Phi(Q^2),
\end{eqnarray}
%%%%%%%%%
where
%%%%%%%%%%%%
\begin{eqnarray}
\label{Eq13}
\Phi(Q^2) &=& \int_{\varepsilon_1}^{\varepsilon_2}[I_{\nu}(\varepsilon_i)+
I_{\bar{\nu}}(\varepsilon)]d\varepsilon_i
\end{eqnarray}
%%%%%%%%%
is the neutrino and antineutrino flux which gives the contribution to the 
measured cross section at the fixed value of $Q^2$. The flux is a function of 
$Q^2$ because $\varepsilon_1=\varepsilon_{min}(Q^2)$ and 
$\varepsilon_2=\varepsilon_{max}$, where $\varepsilon_{max}$ is the maximal 
energy in the (anti)neutrino spectrum. The limit $\varepsilon_{min}(Q^2)$, and 
ultimately the flux $\Phi(Q^2)$ depend on nuclear model. As a result the 
extracted flux-averaged cross section $\langle d\sigma/dQ^2 \rangle$ is 
model-dependent too.   
 
In Eq.\eqref{Eq11} the cross section $d\sigma/d Q^2$ is defined as
%%%%%%%%%%%%
\begin{eqnarray}
\label{Eq14}
\frac{d\sigma}{dQ^2}(Q^2,T_{th},\varepsilon_i) &=& 
\int_{\omega_{min}}^{\omega_{cut}}\frac{d^2\sigma}{dQ^2d\omega}(Q^2, \omega)d\omega,
\end{eqnarray}
%%%%%%%%%
where $\omega_{cut}=min\{\omega_{max}(Q^2),\varepsilon_i-m_{\mu}-T_{th}\}$, 
$m_{\mu}$ is the muon mass, $\omega_{max}(Q^2)$ and $\omega_{min}(Q^2)$ are the 
limits of the kinematic allowed $\omega$-range at the fixed value of $Q^2$. 
If $T_{th}=0$ the upper limit $\omega_{cut}=\omega_{max}(Q^2)$. So, the 
flux-averaged differential $\langle d\sigma^2/dT d\cos\theta\rangle$ and 
$\langle d\sigma/dQ^2\rangle$ cross sections are model-dependent.

In Ref.~\cite{MiniB2} the differential cross sections were extracted using the 
flux $\Phi_{BNB}$ that was determined by integration the 
Booster Neutrino Beamline flux~\cite{Flux} over 
$0\leq \varepsilon_i \leq 3$ GeV, i.e. $\Phi_{BNB}$ is a single number 
($2.90 \times 10^{11}~ \nu_{\mu}/cm^2$). Therefore, 
these flux-integrated differential cross sections are not model-dependent and 
can be written as follows
\begin{eqnarray}
\label{Eq15}
\left( \frac{d^2\sigma}{dTd\cos\theta}\right)^{int} &=& 
\left(\frac{d^2\sigma^{\nu}}{dTd\cos\theta}\right)^{int} +
\left(\frac{d^2\sigma^{\bar{\nu}}}{dTd\cos\theta}\right)^{int},
\end{eqnarray}
where
%%%%%%%%%%%%
\begin{eqnarray}
\label{Eq16}
\left(\frac{d^2\sigma^{\nu,~\bar{\nu}}}{dTd\cos\theta}(T,\cos\theta)
\right)^{int} &=& 
\int_{\varepsilon_1}^{\varepsilon_2}\widetilde{W}_{\nu,~\bar{\nu}}
(T,\cos\theta,\varepsilon_i)\frac{d^2\sigma^{\nu,~\bar{\nu}}}
{dTd\cos\theta}(T,\cos\theta,\varepsilon_i) d\varepsilon_i,
\end{eqnarray}
%%%%%%%%%
and  
\begin{eqnarray}
\label{Eq17}
\left(\frac{d\sigma}{dQ^2}\right)^{int} &=& 
\left(\frac{d\sigma^{\nu}}{dQ^2}\right)^{int} +
\left(\frac{d\sigma^{\bar{\nu}}}{dQ^2}\right)^{int},
\end{eqnarray}
where
%%%%%%%%%%%%
\begin{eqnarray}
\label{Eq18}
\left(\frac{d\sigma^{\nu,~\bar{\nu}}}{dQ^2}(Q^2,T_{th})\right)^{int} &=& 
\int_{\varepsilon_1}^{\varepsilon_2}\widetilde{W}_{\nu,~\bar{\nu}}(Q^2,\varepsilon_i)
\frac{d\sigma^{\nu,~\bar{\nu}}}{dQ^2}(Q^2,T_{th},\varepsilon_i) 
d\varepsilon_i.
\end{eqnarray}
%%%%%%%%%
% FIGURE 2
\begin{figure*}
  \begin{center}
    \includegraphics[height=16cm,width=16cm]{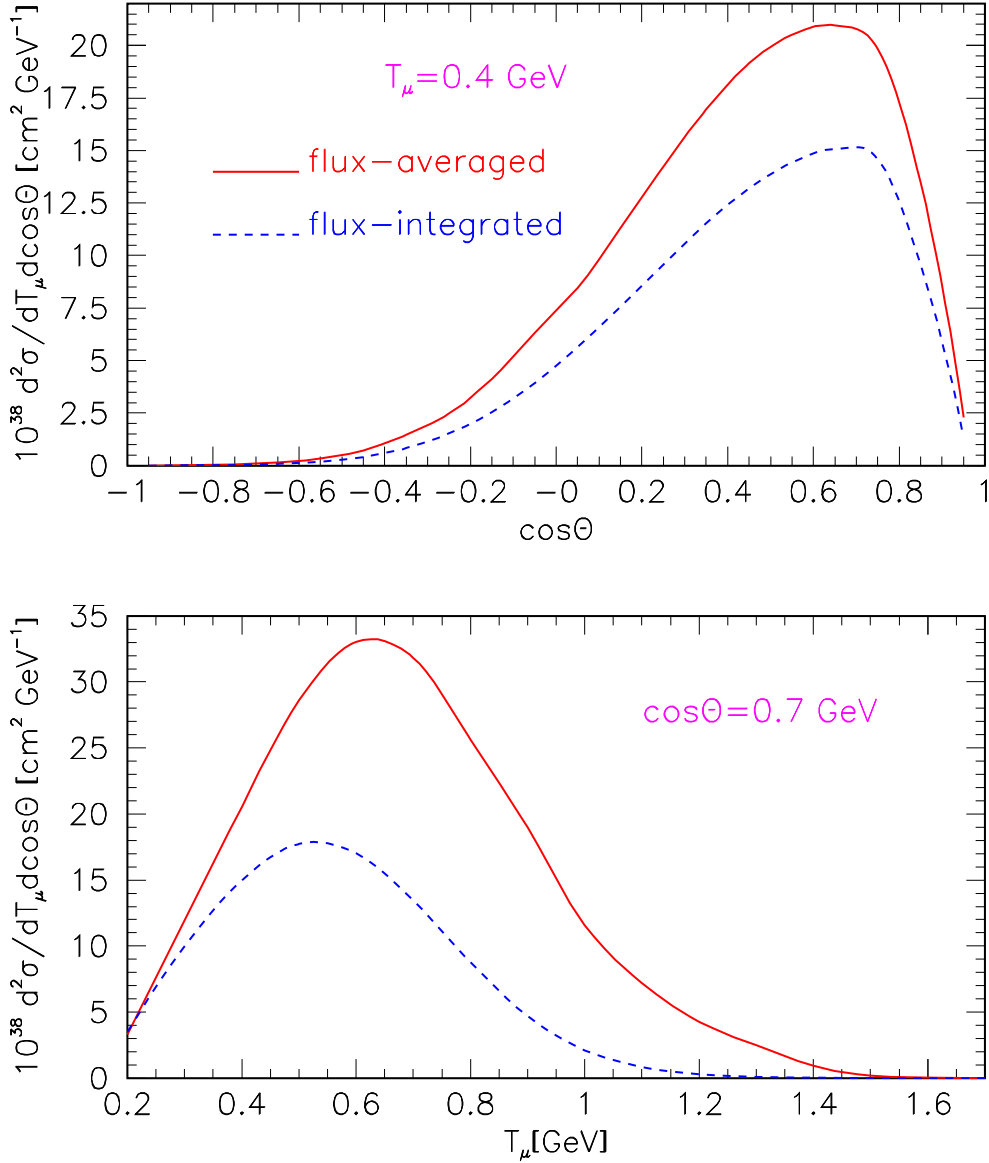}
  \end{center}
  \caption{\label{Fig2}(Color online) Flux-averaged (solid line) and 
flux-integrated (dashed line) double differential cross sections versus 
$\cos\theta$ for $T=0.4$ GeV (upper panel) and versus $T$ for $\cos\theta=0.7$
 (lower panel) calculated in the RDWA approach for $\nu$- mode of the BNB 
flux.} 
\end{figure*}
%%%%%%%%%%%%%%%%%%%%%%%%%%%%%%%%%%%%
The weight functions $\widetilde{W}_{\nu,~\bar{\nu}}$ are defined as
%%%%%%%%%%%%
\begin{eqnarray}
\label{Eq19}
\widetilde{W}_{\nu,~\bar{\nu}}(T,\cos\theta,\varepsilon_i) &=& 
I_{\nu,~\bar{\nu}}(\varepsilon_i)/\Phi_{BNB},
\end{eqnarray}
%%%%%%%%%
and
%%%%%%%%%%%%
\begin{eqnarray}
\label{Eq20}
\int_{\varepsilon_1}^{\varepsilon_2}[\widetilde{W}_{\nu}(T,\cos\theta,\varepsilon_i) + 
\widetilde{W}_{\bar{\nu}}(T,\cos\theta,\varepsilon_i)]d\varepsilon_i &\leq& 1.
\end{eqnarray}
%%%%%%%%%
because of $\Phi_{BNB}\geq\Phi(T,\cos\theta)$ and $\Phi_{BNB}\geq\Phi(Q^2)$.
These functions depend only on (anti)neutrino energy and are model-independent.
As an example, on Fig.~\ref{Fig2} the flux-averaged and flux-integrated double 
differential cross sections calculated within the RDWIA model for $\nu$- mode 
of the BNB flux are compared. Apparently the flux-averaged cross sections are 
higher than the flux-integrated ones. This is because the normalization of 
$\widetilde{W}_{\nu,~\bar{\nu}}$ (Eq.\eqref{Eq20}) is less than unit. 
From the practical point of view, the flux-integrated differential cross 
sections are more useful than flux-averaged ones because they are not 
model-dependent and can be used for comparison to models of CCQE interaction on
 nuclear targets.    

\section{Results and analysis}

\subsection{CCQE flux-integrated $d\sigma/dQ^2$ differential cross section}

New data for CCQE events $Q^2$-distribution measured in the MiniBooNE 
experiment were presented in Refs.~\cite{MiniB2,MiniB3}. The CC one pion 
production (CC1$\pi^+$) background was measured and subtracted instead of 
calculated one~\cite{MiniB1}. With measured CC1$\pi^+$ background 
incorporated, a ``shape-only'' fit to the CCQE events sample was performed to 
extract values for adjusted CCQE model parameters, $M_A$ and $\kappa$ within 
the Fermi gas model. To tune this model to the low $Q^2$, the parameter 
$\kappa$~ was introduced~\cite{MiniB1} which reduced the phase space volume at 
low-momentum transfer. Note that at $\kappa=1$ the phase space volume is the 
same as well as in the ``standard'' RFGM. This parameter controls the $Q^2$-
distribution only in the low-$Q^2$ region. The shape-only fit yields the model 
parameters, $M_A=1.35\pm 0.17 GeV/c^2$ and $\kappa=1.007\pm 0.012$. The 
extracted value for $M_A$ is approximately 30\% higher than the world averaged 
one.      

The MiniBooNE $\nu_{\mu}$ CC flux-integrated single differential cross section 
$d\sigma/dQ^2$ per neutron was extracted as a function of $Q^2$ in the range 
$0 \leq Q^2\leq 2$ (GeV/c)$^2$. To extract value for the parameter $M_A$ we 
calculated this cross section with the BNB flux in the RDWIA and RFGM models 
using the $Q^2$-bins $\Delta Q^2=Q^2_{i+1}-Q^2_i$ similar to Ref.~\cite{MiniB2}
%%%%%%%%%%%%
\begin{eqnarray}
\label{Eq21}
\left(\frac{d\sigma}{dQ^2}\right)^{int}_i=
\frac{1}{\Delta Q^2}\int_{Q^2_i}^{Q^2_{i+1}}
\left[\frac{d\sigma}{dQ^2}(Q^2)\right]^{int}dQ^2 
\end{eqnarray}
%%%%%%%%%
Because the data include events with $T_{\mu}\leq 200$ MeV~\cite{MiniB2}, 
we calculated $d\sigma/d Q^2$ with $T_{\mu}=0$ in Eq.~\eqref{Eq14}. 
%%%%%%%%%%%%%%%%%%%%%        
% FIGURE 3
\begin{figure*}
  \begin{center}
    \includegraphics[height=16cm,width=16cm]{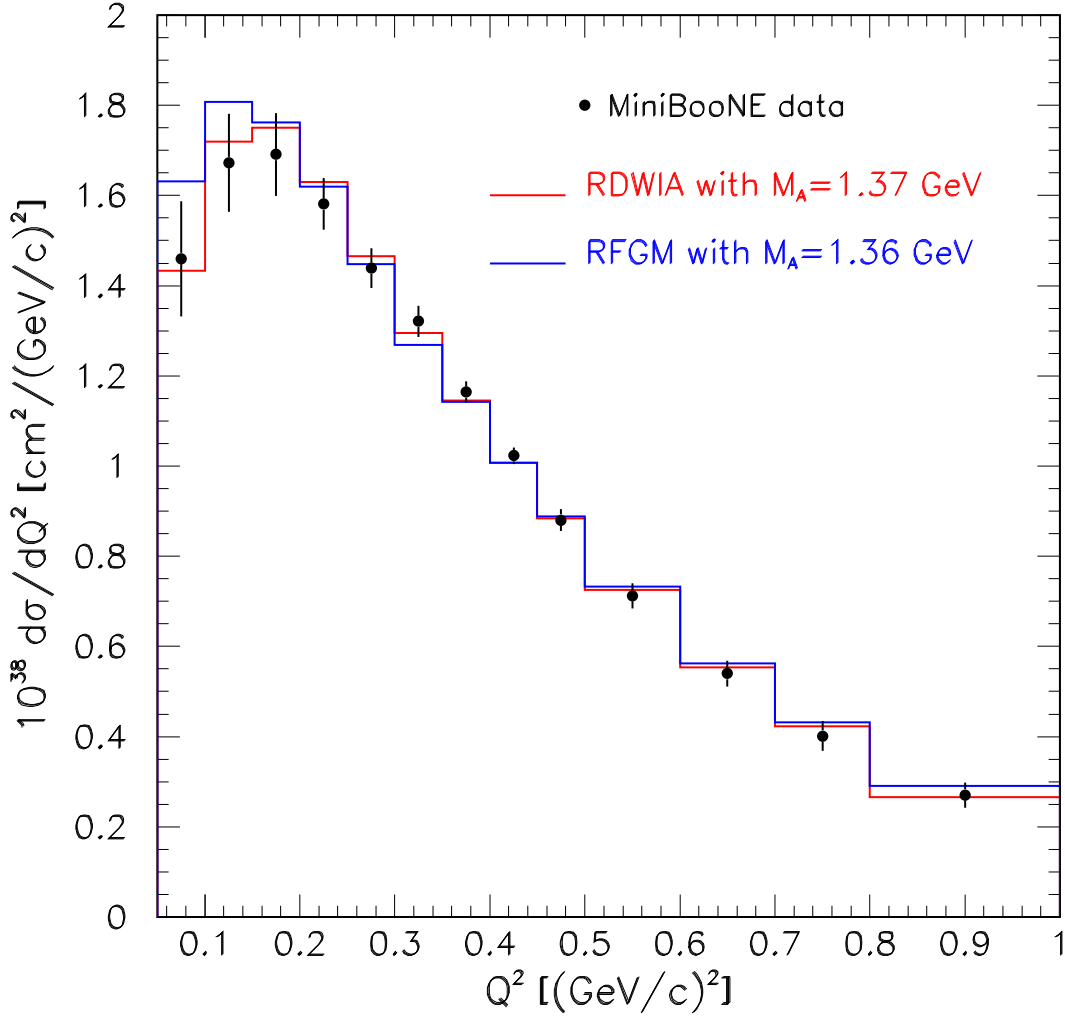}
  \end{center}
  \caption{\label{Fig3}(Color online) Flux-integrated $d\sigma/dQ^2$ cross 
section per neutron target for the $\nu_{\mu}$ CCQE scattering. Calculations 
from the RDWIA with $M_A=1.37$ GeV/c$^2$ and RFGM with $M_A=1.36$ GeV/c$^2$. 
The MiniBooNE data are shown as points with the shape error only.} 
\end{figure*}
%%%%%%%%%%%%%%%%%%%%%%%%%%%%%%%%%%%%
Within the RDWIA (RFGM) model the fit to the extracted 
flux-integrated $d\sigma/dQ^2$ cross section with only-shape error yields the 
parameter $M_a=1.37\pm 0.05$ GeV/c$^2$ $(M_A=1.36\pm 0.05$ GeV/c$^2)$. These 
values are consistent with the MiniBooNE result $M_A=1.37\pm 0.17$ GeV/c$^2$.

Figure 3 shows measured flux-integrated $d\sigma/dQ^2$ differential cross 
section as a function of $Q^2$ compared with the RDWIA ($M_A=1.37$ GeV/c$^2$) 
and RFGM ($M_A=1.36$ GeV/c$^2$) calculations. There is an overall agreement 
between the RDWIA result and the data across of the full range $Q^2=0\pm 1$ 
(GeV/c)$^2$, whereas the RFGM overestimates the measured differential cross 
section at $Q^2\leq 0.2$ (GeV/c)$^2$. At higher $Q^2$ a good match between the 
RFGM calculated and measured cross sections is observed. Thus, so-called 
low-$Q^2$ problem is successfully solved in the distorted-wave approach. 
 
\subsection{CCQE flux-integrated double differential cross section}

The flux-integrated double differential cross section per neutron 
$d^2\sigma/dTd\cos\theta$, for the $\nu_{\mu}$ CCQE process was extracted in 
Ref.~\cite{MiniB2} for the kinematic range, $-1 < \cos\theta < 1$, 
$0.2 < T < 2$ GeV. The flux-integrated CCQE total cross section, obtained by 
integrating the double differential one over this range was measured to be 
$18.447\times10^{-39}$ cm$^2$ and $9.429\times 10^{-39}$ cm$^2$ for range 
$-1 < \cos\theta < 1$, $0 < T < 2$ GeV. The total normalization error on this 
measurement is 10.7\%. These results contain the most complete and 
model-independent information that is available from experiment on the 
CCQE process. 
 
We calculated the flux-integrated double differential cross section 
$\left(d^2\sigma/dT d\cos\theta)^{int}\right)$ for the BNB $\nu_{mu}$ flux 
within the RDWIA and RFGM models with the extracted values of $M_A$ using 
the $T$ and $Q^2$-bins similar to Ref.~\cite{MiniB2}
%%%%%%%%%%%%
\begin{eqnarray}
\label{Eq22}
\left(\frac{d^2\sigma}{dT d\cos\theta}\right)^{int}_{ij}=
\frac{1}{\Delta T \Delta\cos\theta}\int_{T_i}^{T_{i+1}}\int_{(\cos\theta)_j}
^{(\cos\theta)_{j+1}}
\left[\frac{d^2\sigma}{dT d\cos\theta}(T,\cos\theta)\right]^{int}dT d\cos\theta, 
\end{eqnarray}
%%%%%%%%%
where $\Delta T=T_{i+1}-T_i=0.1$ GeV and $\Delta\cos\theta=(\cos\theta)_{j+1}-
(\cos\theta)_j=0.1$.

\begin{figure*}
  \begin{center}
    \includegraphics[height=17cm,width=17cm]{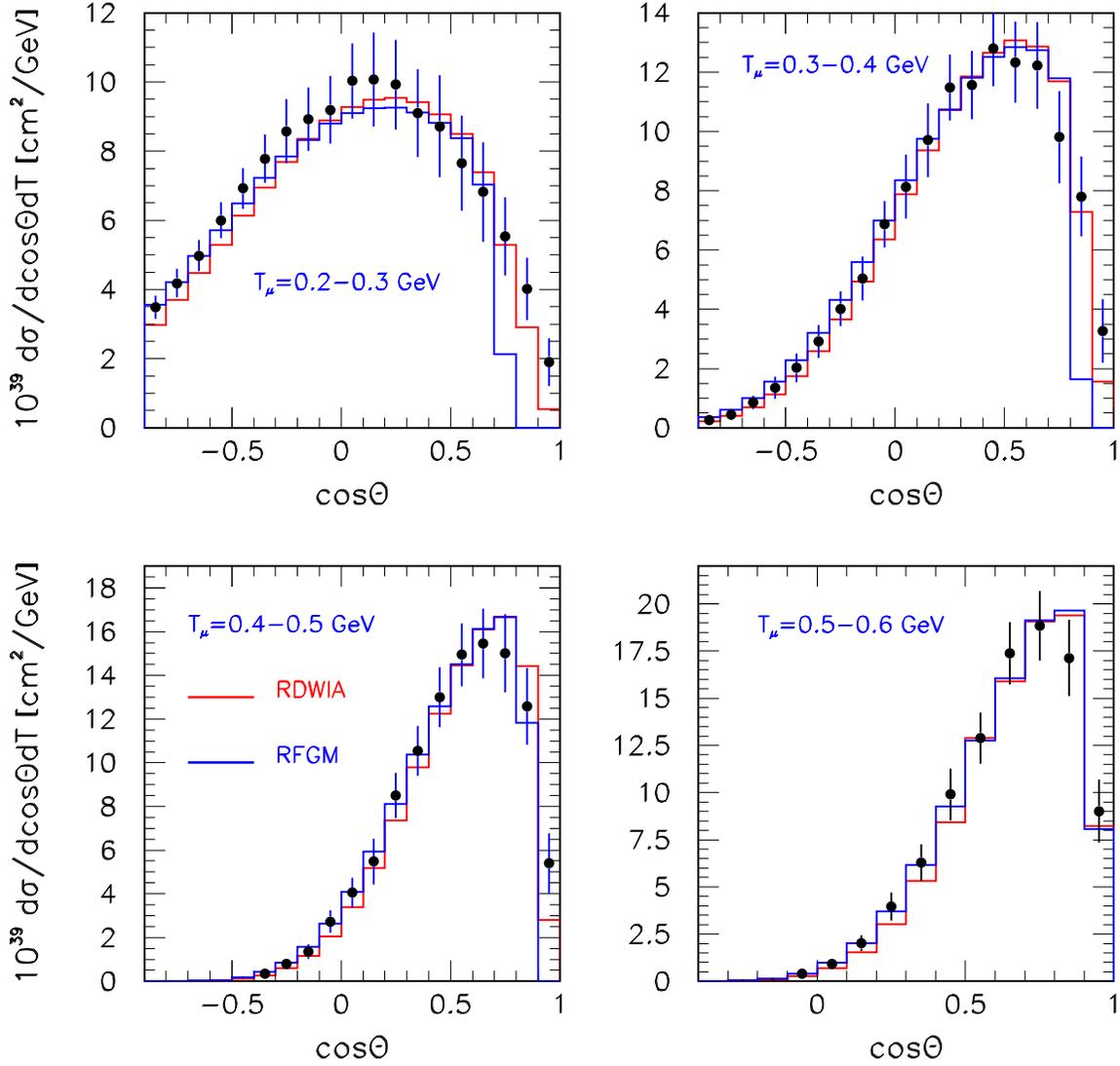}
  \end{center}
  \caption{\label{Fig4}(Color online) Flux-integrated 
$d^2\sigma/dT d\cos\theta$ cross section per neutron target for the 
$\nu_{\mu}$ CCQE process as a function of $\cos\theta$ for the four muon kinetic
 energy bins: $T$(GeV)=(0.2 - 0.3), (0.3 - 0.4), (0.4 - 0.5), and (0.5 - 0.6). 
As shown in the key, cross sections were calculated within the RDWIA 
($M_A=1.37$ GeV/c$^2$) and RFGM ($M_A=1.36$ GeV/c$^2$). 
The MiniBooNE data are shown as points with the shape error only.} 
\end{figure*}

\begin{figure*}
  \begin{center}
    \includegraphics[height=17cm,width=17cm]{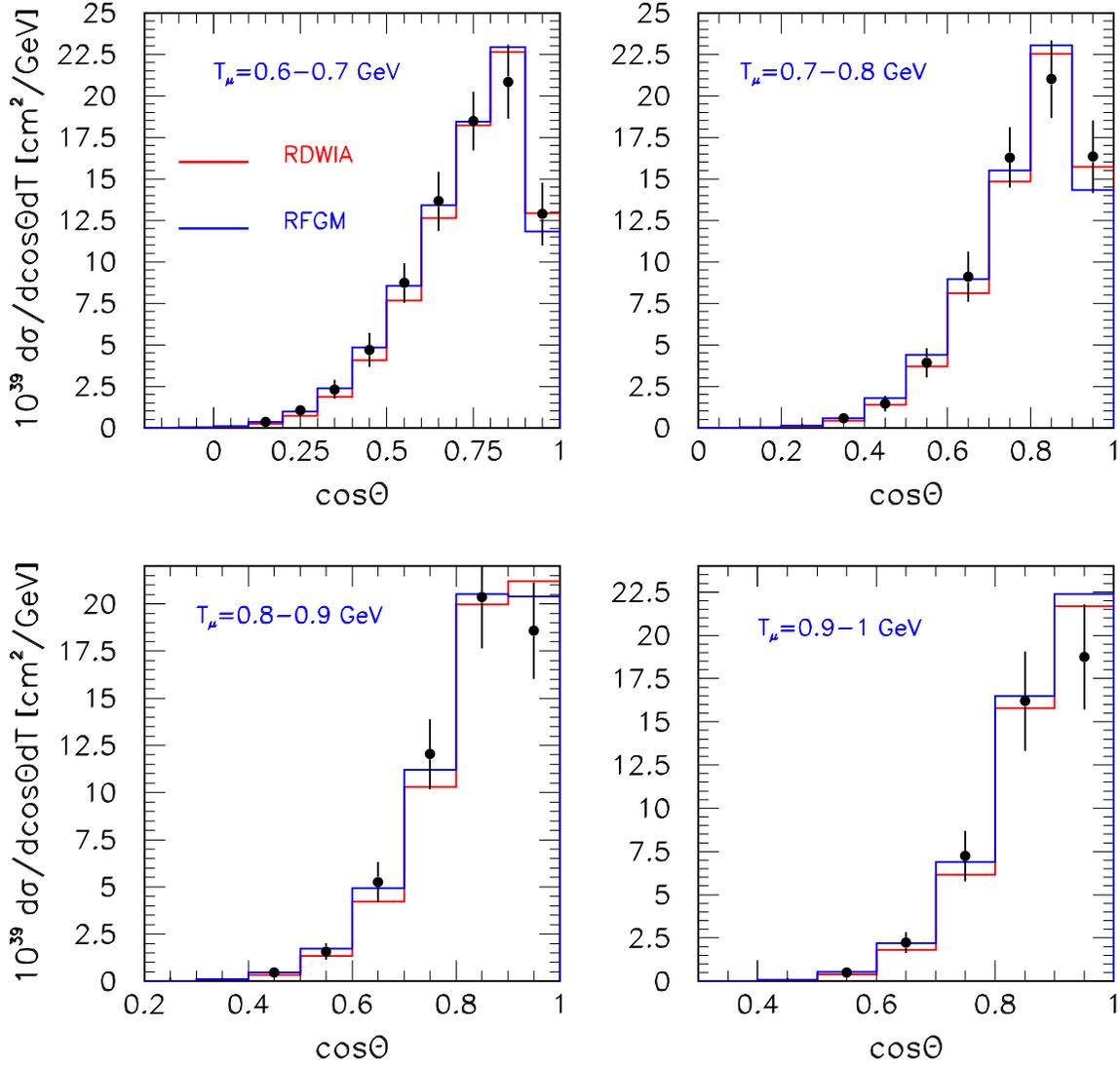}
  \end{center}
  \caption{\label{Fig5}(Color online) Same as Fig.~\ref{Fig4} but for muon 
kinetic energy bins: $T$(GeV)=(0.6 - 0.7), (0.7 - 0.8), (0.8 - 0.9), and 
(0.9 - 1).} 
\end{figure*}

Figures~\ref{Fig4} and~\ref{Fig5} show measured flux-integrated 
$d^2\sigma/dTd\cos\theta$ cross sections as functions of $\cos\theta$ for 
several bins of muon kinetic energy in the range $0.2\leq T \leq 2$ GeV as 
compared with the RDWIA and RFGM calculations. There is good agreement between 
the RDWIA calculations and data within the error of the experiment. But in 
the regions $0.2\leq T \leq 0.3$ GeV, $-1\leq \cos\theta \leq -0.3$ and 
$0.2\leq T \leq 0.5$ GeV, $0.9 \leq \cos\theta \leq 1$ the RDWIA 
result are slightly lower then measured cross section and the difference 
decreases with muon energy.  
%%%%%%%%%%%%%%%%%%%%%        
% FIGURE 4
%%%%%%%%%%%%%%%%%%%%%%%%%%%%%%%%%%%%
%%%%%%%%%%%%%%%%%%%%%        
% FIGURE 5
%%%%%%%%%%%%%%%%%%%%%%%%%%%%%%%%%%%%
%%%%%%%%%%%%%%%%%%%%%        
% FIGURE 6
\begin{figure*}
  \begin{center}
    \includegraphics[height=17cm,width=17cm]{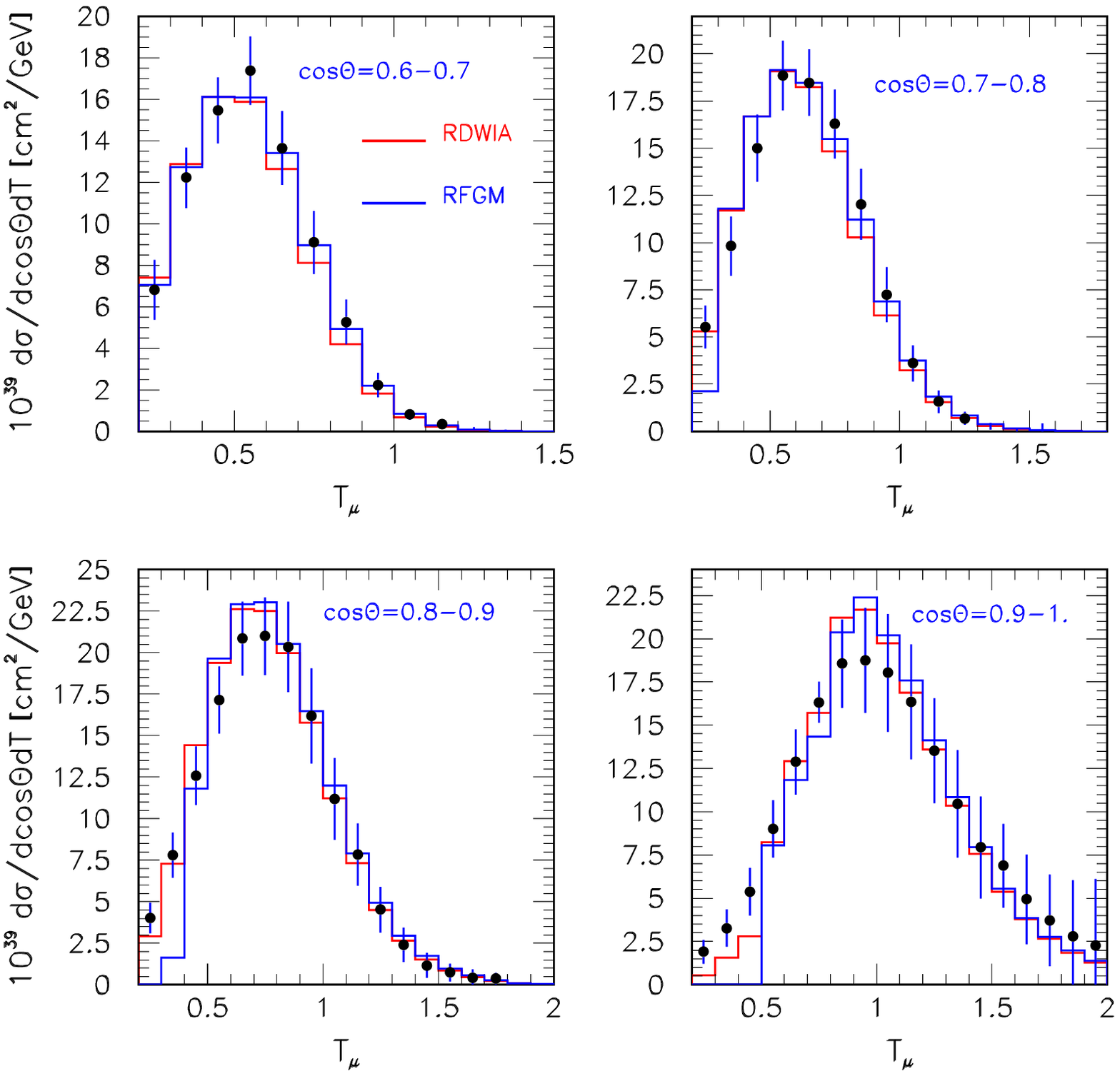}
  \end{center}
  \caption{\label{Fig6}(Color online) Flux-integrated 
$d^2\sigma/dT d\cos\theta$ cross section per neutron target for the 
$\nu_{\mu}$ CCQE process as a function of muon energy for the four muon 
scattering angle bins: $\cos\theta$=(0.6 - 0.7), (0.7 - 0.8), (0.8 - 0.9), 
and (0.9 - 1). As shown in the key, cross sections were calculated within the 
RDWIA ($M_A=1.37$ GeV/c$^2$) and RFGM ($M_A=1.36$ GeV/c$^2$). 
The MiniBooNE data are shown as points with the shape error only.} 
\end{figure*}
%%%%%%%%%%%%%%%%%%%%%%%%%%%%%%%%%%%%

The RFGM prediction also agree well with data within the errors, except the 
region $0.7 < \cos\theta < 1$ and $0.2 < T <0.5$ GeV, where the calculated 
cross sections fall down rapidly with $\cos\theta$. In this kinematic region 
the Fermi gas model underestimates the double differential cross section 
significantly. This trend is characteristic of nucleon momentum distribution 
and Pauli bloking effect as calculated in the Fermi gas model~\cite{BAV4}.

Fifure~\ref{Fig6} shows measured flux-integrated $d^2\sigma/dT d\cos\theta$ 
cross sections as functions of muon energy for four bins of muon scattering 
angle as compared with the RDWIA and RFGM calculations. Apparently that the 
RDWIA cross sections are lower than the measured ones in the kinematic region 
$0.9 < \cos\theta < 1$, $0.2 \leq T \leq 0.5$ (GeV) and the RFGM calculation 
underestimates the measured double differential cross section significantly in 
the range $0.7 < \cos\theta < 1$, $0.2 < T < 0.5$ GeV.

So, the comparison measured and calculated flux-integrated 
$d^2\sigma/dT d\cos\theta$ cross sections shows that the Fermi gas model 
prediction are completely off the data in the range $0.7 < \cos\theta < 1$, 
$0.2 < T < 0.5$ GeV. The RDWIA cross sections underestimate the measured ones 
for muon production with energies $T \leq 0.3$ GeV and scattering angles 
$\cos\theta > 0.9$.

\subsection{CCQE flux-integrated $d\sigma/dT$ and $d\sigma/d\cos\theta$ cross 
section}

\begin{figure*}
  \begin{center}
    \includegraphics[height=17cm,width=17cm]{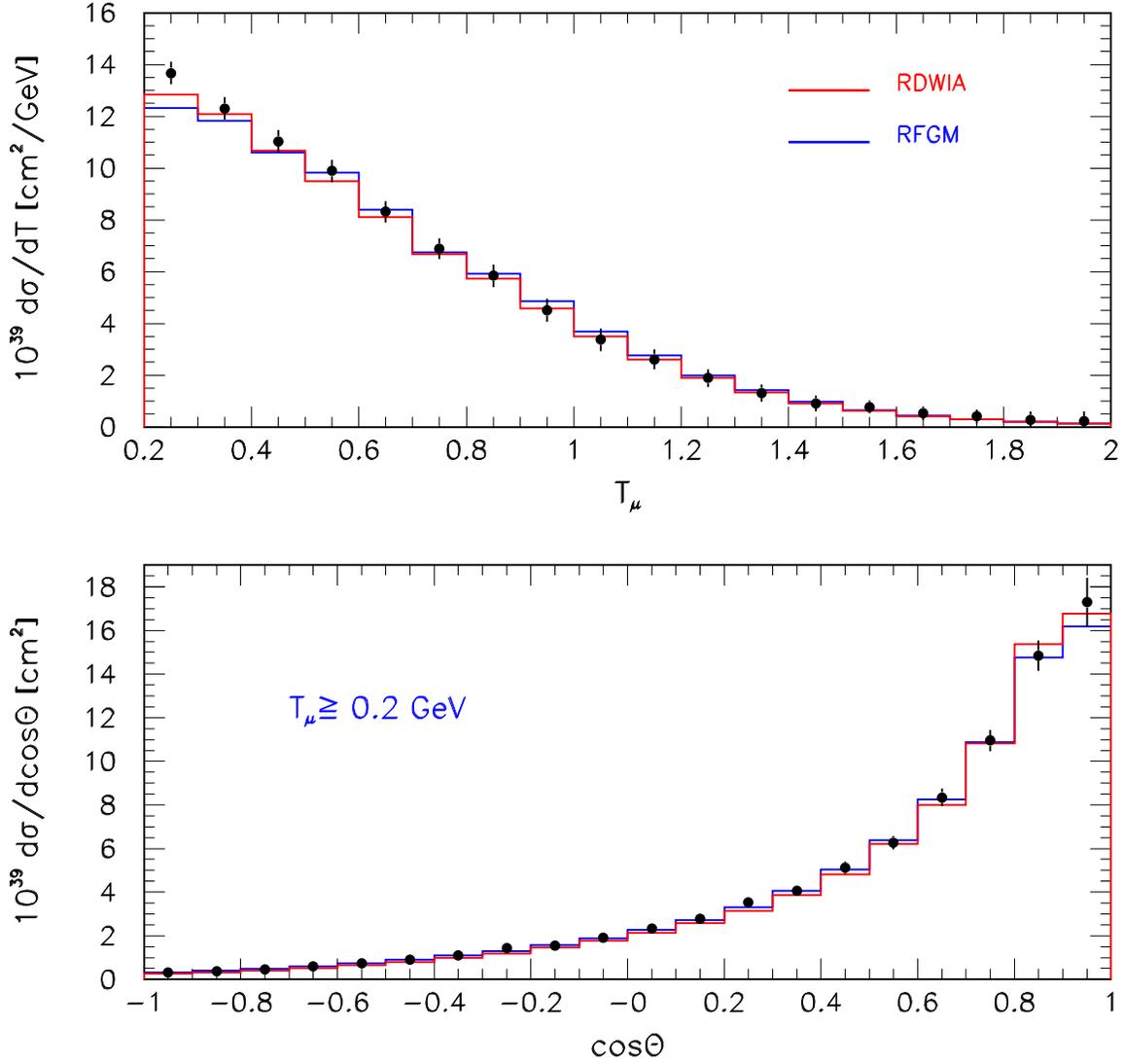}
  \end{center}
  \caption{\label{Fig7}(Color online) Flux-integrated 
$d\sigma/dT$ cross section as a function of muon energy (upper panel) and 
 $d\sigma/d\cos\theta$ cross section for $T\geq 0.2$ GeV as a function of muon 
scattering angle (lower panel) for the $\nu_{\mu}$ CCQE process. 
As shown in the key, cross sections were calculated within the RDWIA and RFGM. 
The MiniBooNE data are shown as points with the shape error only.} 
\end{figure*}

The flux-integrated single differential cross sections $d\sigma/dT$ and 
$d\sigma/d\cos\theta$ (for $T\geq 0.2$ GeV) are presented in Fig.~\ref{Fig7}, 
which shows $d\sigma/dT$ as a function of kinetic muon energy and 
$d\sigma/d\cos\theta$ as a function of muon scattering angle. Here the results 
obtained in the RDWIA and Fermi gas models compared with the MiniBooNE data. 
The measured flux-integrated $d\sigma/dT$ ($d\sigma/d\cos\theta$) cross 
section with the shape error has been obtained by summing the double 
differential one over $\cos\theta$-bins ($T$-bins) presented in Tables VI 
and VII in Ref.~\cite{MiniB2}. There is a good agreement between the 
calculated and measured cross sections, with the excepton of the bin 
$0.2 \leq T \leq 0.3$ GeV. The flux integrated total cross sections obtained 
in the RDWIA and RFGM approaches by integrating the double differential cross 
sections (over $-1 \leq \cos\theta \leq 1$, $0.2 \leq T \leq 2$ GeV), 
are equal of $8.208\times 10^{-39}$ cm$^2$ and $8.310\times 10^{-39}$ cm$^2$, 
correspondingly, and agree with measured one of $8.447\times 10^{-39}$ cm$^2$. 
%%%%%%%%%%%%%%%%%%%%%        
% FIGURE 7
%%%%%%%%%%%%%%%%%%%%%%%%%%%%%%%%%%%%
%%%%%%%%%%%%%%%%%%%%%        
% FIGURE 8
\begin{figure*}
  \begin{center}
    \includegraphics[height=17cm,width=17cm]{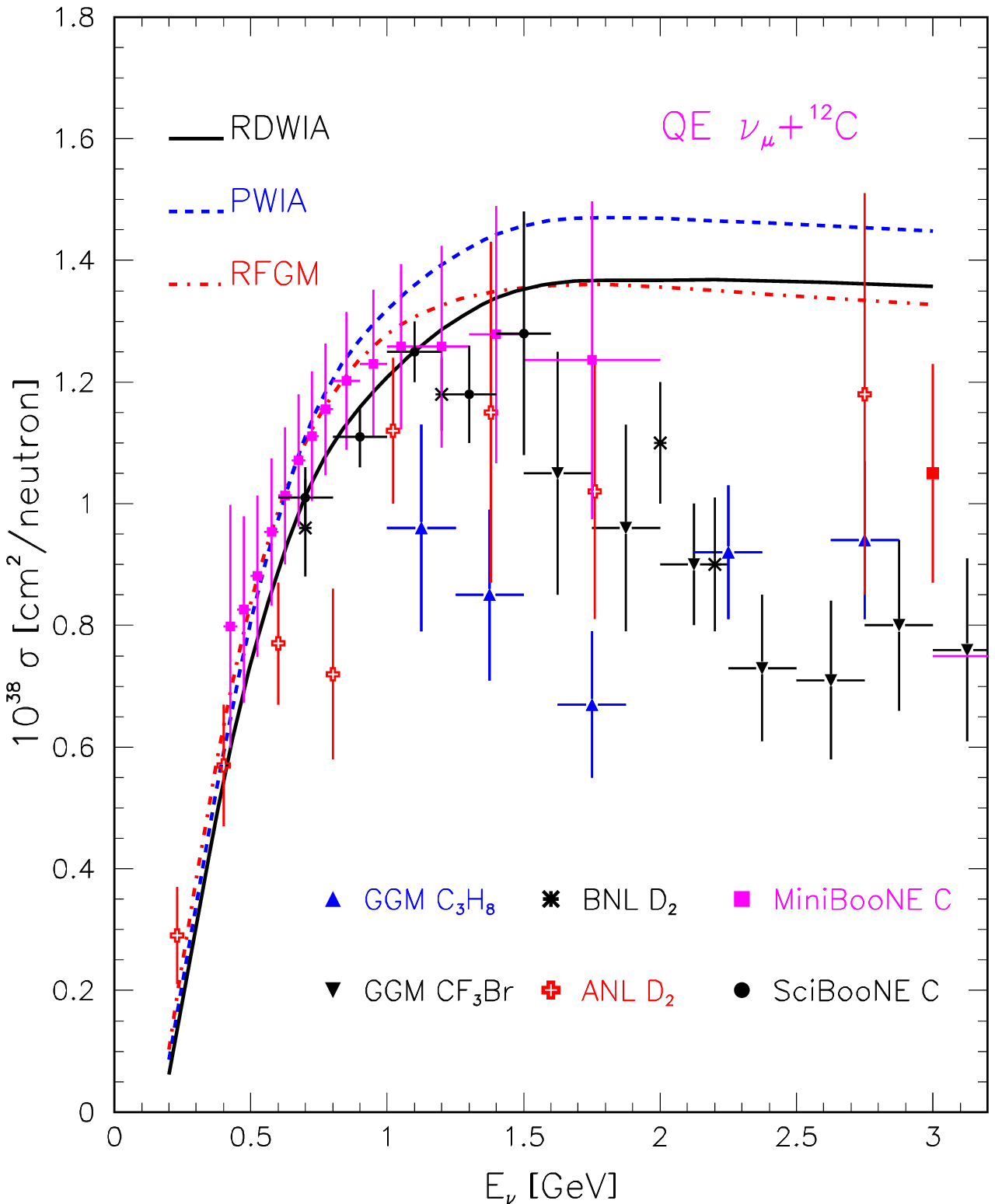}
  \end{center}
  \caption{\label{Fig8}(Color online) Total $\nu_{\mu}$ CCQE cross section 
per neutron as a function of neutrino energy. Data points for different targets
 are from~\cite{SciB,MiniB2,Mann,Baker,Pohl,Brunner}. Also shown are 
predictions of the RDWIA ($M_A=1.37$ GeV/c$^2$), PWIA ($M_A=1.37$ GeV/c$^2$),
 and RFGM ($M_A=1.36$ GeV/c$^2$).}
\end{figure*}
%%%%%%%%%%%%%%%%%%%%%%%%%%%%%%%%%%%%

\subsection{CCQE total cross section}

The MiniBooNE flux-unfolded CCQE cross section per neutron as a function of 
neutrino energy is shown in Fig.~\ref{Fig8} together with data of 
Refs.~\cite{Mann, Baker, Pohl, Brunner}. Also shown for comparison are the 
results obtained in the RDWIA, PWIA, and RFGM approaches. The calculated cross 
sections with the values of $M_A$, extracted from the shape-only fit to the 
flux-integrated $d\sigma/Q^2$ data reproduce the MiniBooNE total cross section 
within the errors of experiment over the entire measured energy range. At the 
average energy of the MiniBooNE flux ($\approx 800$ MeV), the extracted cross 
section is $\approx 30\%$ higher than 
what is commonly assumed for this process assuming the RFGM and world-average 
value of the axial mass, $M_A=1.03$ GeV/c$^2$.
Note, that the spread in the data is much higher than a difference in 
predictions of the RDWIA, PWIA, and RFGM approaches. So, the comparison of the 
predicted and measured model-independent flux-integrated double differential 
cross sections is more sensitive test of the employed models of the CCQE 
process than the comparison of the total cross sections.  

\section{Conclusions}

In this paper, we analyze the flux-averaged and flux-integrated differential and
 total $\nu_{\mu}$ CCQE cross sections placing particular emphasis on their 
nuclear-model dependence. We found that the flux-integrated cross sections are 
model-independent and can be used to test of employed models of the CCQE 
interaction on nuclear targets. The flux-integrated double differential 
$d^2\sigma/dTd\cos\theta$, single differential $d\sigma/dQ^2$, $d\sigma/dT$, 
$d\sigma/d\cos\theta$, and flux-unfolded $\sigma(\varepsilon_i)$ CCQE cross 
sections were measured in the MiniBooNE experiment~\cite{MiniB2}.

Using the RDWIA and RFGM approaches with the BNB flux we extracted an axial 
mass from a ``shape-only'' fit of the measured flux-integrated $d\sigma/dQ^2$ 
differential cross section. The extracted value of 
$M_A=1.37\pm 0.05$ GeV/c$^2$ (RDWIA) and $M_A=1.36\pm 0.05$ GeV/c$^2$ (RFGM) 
that is consistent with the MiniBooNE result of $M_A=1.35\pm 0.17$ GeV/c$^2$. 
The flux-integrated double differential cross sections were calculated in 
these models with extracted values of $M_A$. There is an overall agreement 
between the RDWIA result and data, whereas the RFGM calculation overestimates 
the measured cross section at $Q^2 < 0.2$ (GeV/c)$^2$. Thus, so-called 
low-$Q^2$ problem is successfully solved in the framework of RDWIA. 

We also calculated in the RDWIA and RFGM approaches the flux-integrated    
$d^2\sigma/dTd\cos\theta$, $d\sigma/dQ^2$, $d\sigma/dT$ (for muons with kinetic
 energy $T\geq 0.2$ GeV), and total cross sections and compared them with the 
MiniBooNE data. The comparison of the RDWIA double differential cross section 
shows good agreement with data within the error of the experiment, except in 
the region $0.2 \leq T \leq 0.3$ GeV, $0.9 \leq\cos\theta\ \leq 1$ where the 
calculated cross sections are lower then measured ones. A good agreement between
 the RFGM calculation and data is observed exclusive of the range 
$0.7\leq \cos\theta 1$, $0.2 \leq T \leq 0.5$ GeV where the Fermi gas model 
predictions are completely off of the data. The calculated $d\sigma/dT$ and 
$d\sigma/d\cos\theta$ also describe well the measured cross sections except the
muon energy bin $0.2 \leq T \leq 0.3$ GeV where the calculations are lower 
then the data.

The calculated and measured flux-integrated total cross sections are match 
well. The RDWIA, PWIA and RFGM calculations with extracted values of $M_A$ 
reproduce the MiniBooNE flux-unfolded CCQE cross section within the 
experimental error over the entire measured energy range.

We conclude that the flux-integrated double differential cross section which 
is model-independent should be used as the preferred choice for comparison to
employed model of the CCQE interaction on nuclear targets.   

\section*{Acknowledgments}

The author greatly acknowledges S. Kulagin, J. Morfin, G. Zeller, and T.Katori 
for fruitful discussions at different stages of this work.  
%

%%%%%%%%%%%%%%%%%%%%%%%%%%%%%%%%%%
%% thebibliography environment %%
%%%%%%%%%%%%%%%%%%%%%%%%%%%%%%%%%

%%%%%%%%%%
\end{document}